\documentclass{aa} 
\bibpunct{(}{)}{;}{a}{}{,} % to follow the A&A style

\usepackage{txfonts}
\usepackage{bm}
\usepackage{ mathrsfs }
\usepackage{courier}
\usepackage{pgfplots}
\usepackage{pgf}
\usepackage{comment}
\usepackage{amsmath}
\usepackage{amssymb}
\usepackage{natbib}
\usepackage{graphicx}
\newcommand{\Al}{$^{26}$Al\ }

\newcommand{\Fe}{$^{60}$Fe\ }

\begin{document} 
   \title{$^{26}$Al gamma rays from the Galaxy with INTEGRAL/SPI}
   \author{  Moritz M. M. Pleintinger \inst{1,2} 
          \and
         Roland Diehl\inst{1,}\thanks{corresponding author, \email{rod@mpe.mpg.de}}
           \and
          Thomas Siegert\inst{1,3}
          \and
          Jochen Greiner\inst{1}
        \and
        Martin G. H. Krause\inst{4}
          }
   \institute{Max-Planck-Institut f\"ur extraterrestrische Physik, Giessenbachstr. 1, D-85748 Garching, Germany
   \and
   Horn \& Company Financial Services GmbH, Kaistr. 20, D-40221 D\"usseldorf, Germany
  \and
	Institut f\"ur Theoretische Physik und Astrophysik, Universit\"at W\"urzburg, Emil-Fischer-Str. 31, D-97074 W\"urzburg, Germany
	   \and
   Centre for Astrophysics Research, School of Physics, Astronomy and Mathematics, University of Hertfordshire, College Lane, Hatfield, Hertfordshire AL10 9AB, UK
  % \and
  %  {corresponding author  \email{rod@mpe.mpg.de}}
}

   \date{Received Sep 27, 2022; revised Dec 16, 2022; accepted Dec 20, 2022}

% \abstract{}{}{}{}{} 
% 5 {} token are mandatory
  \abstract
   {The presence of radioactive $^{26}$Al at 1.8\,MeV reflects ongoing nucleosynthesis in the Milky Way. Diffuse emission from its decay can be measured with gamma-ray telescopes in space. The intensity, line shape, and spatial distribution of the $^{26}$Al emission allow a study of these nucleosynthesis sources. The line parameters trace massive-star feedback in the interstellar medium due to its 1~My lifetime. 
   }
   {We aim to deepen previous studies of the $^{26}$Al emission in the Milky Way, using all gamma-ray data including single and double events as collected with {\rm SPI} on {\it INTEGRAL} from 2003 until 2020. }
   { We apply improved spectral response and background as evaluated from tracing spectral details over the entire mission. The exposure for Galactic $^{26}$Al emission is enhanced using all event types measured within SPI. We re-determine the intensity of Galactic $^{26}$Al emission across the entire sky, through maximum likelihood fits of simulated and model-built sky distributions to SPI spectra for single and for double detector hits. }
   %Interpretations are put in new light using our massive-star group population synthesis tools.}
   {We find an all-sky flux of (1.84$\pm$0.03$)\times$10$^{-3}$~ph~cm$^{-2}$s$^{-1}$ in the 1.809~MeV line from $^{26}$Al, determined as fitted to sky distributions from previous observations with {\rm COMPTEL}. {Significant emission from higher latitudes indicate an origin from  nearby massive-star groups and superbubbles, also supported by a bottom-up population synthesis model.} The line centroid is found at (1809.83$\pm$0.04~keV, and line broadening from source kinematics integrated over the sky is (0.62$\pm0.3$)~keV (FWHM). 
  }
   {}

   \maketitle
%
%--------------------------------------------------------------------------------------------------------

\section{Introduction}
	
   $^{26}$Al has been established as a tracer of ongoing nucleosynthesis in our Galaxy \citep{Prantzos:1996a}. 
   The {\it INTEGRAL} mission  \citep{Winkler:2003} with its  SPI gamma-ray spectrometer \citep{Vedrenne:2003,Roques:2003} has enabled accumulation of observing time since its 2002 launch, continuing to this day, measuring the large-scale Galactic \Al emission. 
   First results \citep{Diehl:2003} were refined to constrain the extent of the Galaxy's \Al emission \citep{Wang:2009}, an all-sky imaging could be obtained \citep{Bouchet:2015} which essentially confirms the COMPTEL result, and improvements in background handling led to another step in precision of SPI's \Al results \citep{Siegert:2016}.   
In this paper we report results from analysis of nearly 18 years of data \citep{Pleintinger:2020}, using both single and double detector events within the SPI camera for this purpose for the first time, while also
	 exploiting the spectral response and background detail that we obtained from a deep study of the behaviour and variations of SPI spectra
	over the entire mission. 

%----------------------------------------------------------------------------------------------------------

\section{Observations and Data Analysis Approach}
\label{sec:observationsAnalysis}

\subsection{INTEGRAL Measurements}
\label{sec:measurements}

The {\it INTEGRAL} space observatory has been launched by ESA in October 2002 into its excentric orbit located outside the radiation belts \citep{Winkler:2003}.
The SPI spectrometer \citep{Vedrenne:2003} is one of the two main instruments on {\it INTEGRAL}. It is specialised to perform high-resolution spectroscopy over the
15~keV to 8~MeV energy range with its 19-element Ge detector camera, with an energy resolution of 3~keV at 1809~keV. 
A coded mask above the camera allows imaging making use 
of detector shadowings by the mask of emission from regions of the sky, to achieve an imaging resolution of about 3~degrees.%$^o$. 
{\it INTEGRAL} is pointed at one target direction in the sky for a duration of approximately 30 minutes, whereafter the pointing direction is normally moved by 2.1$^o$ within a rectangular 5 by 5 pattern around the target region of interest. This "dithering" leads to additional variations of the detector shadowing, as the offset angle of 2.1$^o$ is matched to the geometrical configuration of detector and mask spacings, and corresponds to a one-detector offset of the shadowing for a source on axis.
The sensitivity    (3$\sigma$)  of SPI in the \Al line was estimated to be about 3$\times$10$^{-5}$~ph~cm$^{-2}$s$^{-1}$~per Ms of exposure \citep{Roques:2003}. 
For individual source regions in the plane of the Galaxy a sensitivity (3$\sigma$) of 2.5$\times$10$^{-5}$~ph~cm$^{-2}$s$^{-1}$
has been achieved by now (from a 6~Ms exposure in the example of the Vela region) \citep{Pleintinger:2020}.
 
\begin{figure}[ht] %%%%%%%%%%%%%%%%%%%%%%%%%%%%%%%%%%%%%%%%%
\centering
\includegraphics[width=0.8\columnwidth,clip]{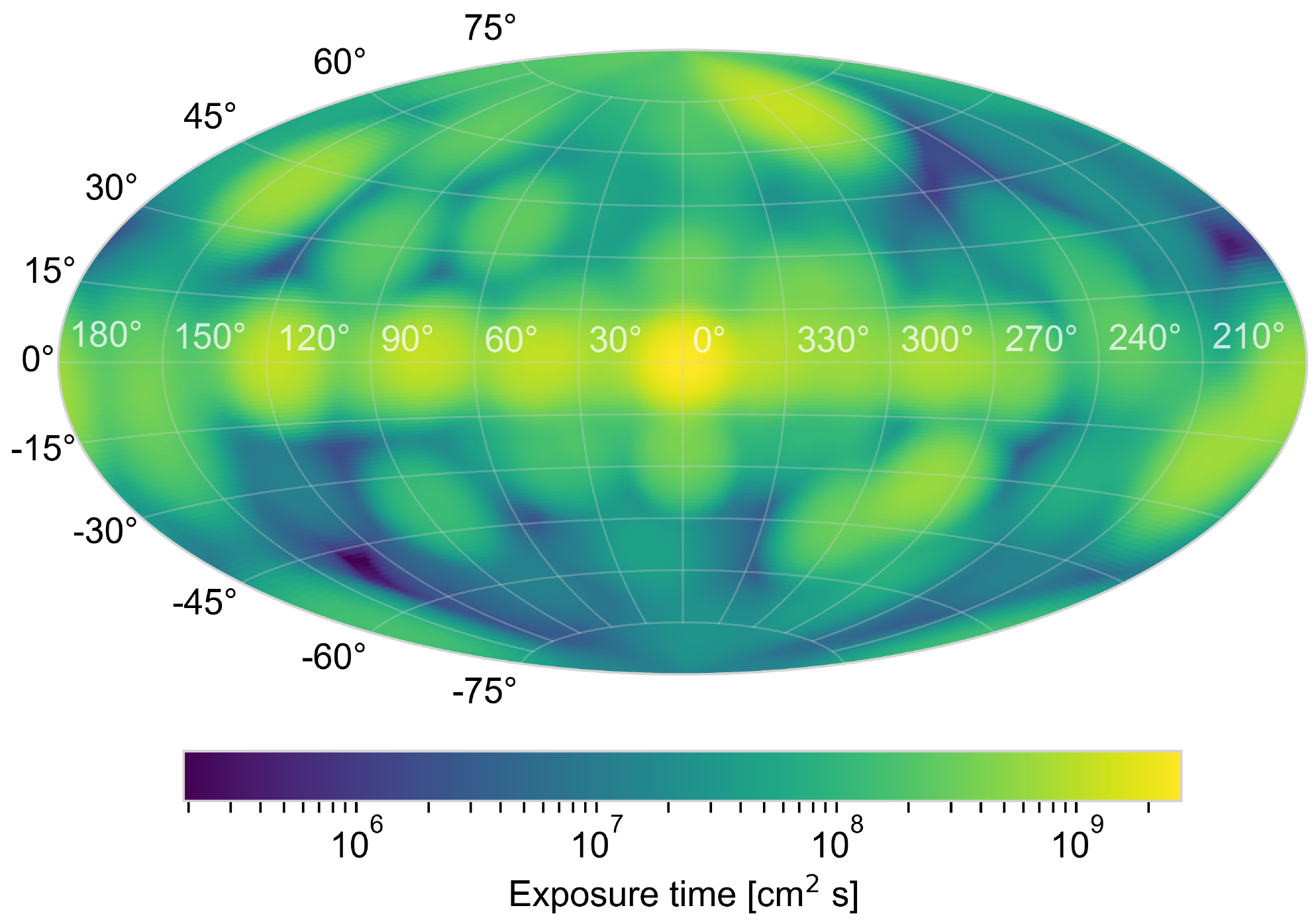}
\caption{The integrated {it INTEGRAL}/SPI exposure of the sky of the data used in this paper. Although {\it INTEGRAL's} emphasis of targets is in the inner Galaxy, large parts of the sky have been exposed to at least 10\% of the Galactic-disk exposure, and allow all-sky analysis.}
\label{fig:exposure}       
\end{figure}

Here we use data from 2002 until 2020, comprising 17.5 years of observations, covering 2131 {\it  INTEGRAL} orbits of typically 3 days duration each.
The achieved sky exposure is illustrated in Fig.~\ref{fig:exposure}, and shows that the full sky is covered, although the {\it INTEGRAL} mission emphasises observations of the inner Galaxy.
The raw data have been filtered to exclude periods where either spacecraft or the SPI instrument were outside their normal conditions, and part of the orbits that might be affected from encounters of the radiation belts near orbit perigee.
The selections involving SPI performance include, in particular, acceptance windows on the rates in the BGO anti-coincidence detector system, and on saturations in the Ge detectors.
After selections, 118407 spacecraft pointings from 1840 orbits remained, providing a sky exposure of about 255~Ms, the distribution of which is shown in Figure~\ref{fig:exposure}.

 %%%%%%%%%%%%%%%%%%%%%%%%%%%%%%%%%%%%%%%%%%%

\begin{figure}[ht] %%%%%%%%%%%%%%%%%%%%%%%%%%%%%%%%%%%%%%%%%
\centering
\includegraphics[width=0.8\columnwidth,clip]{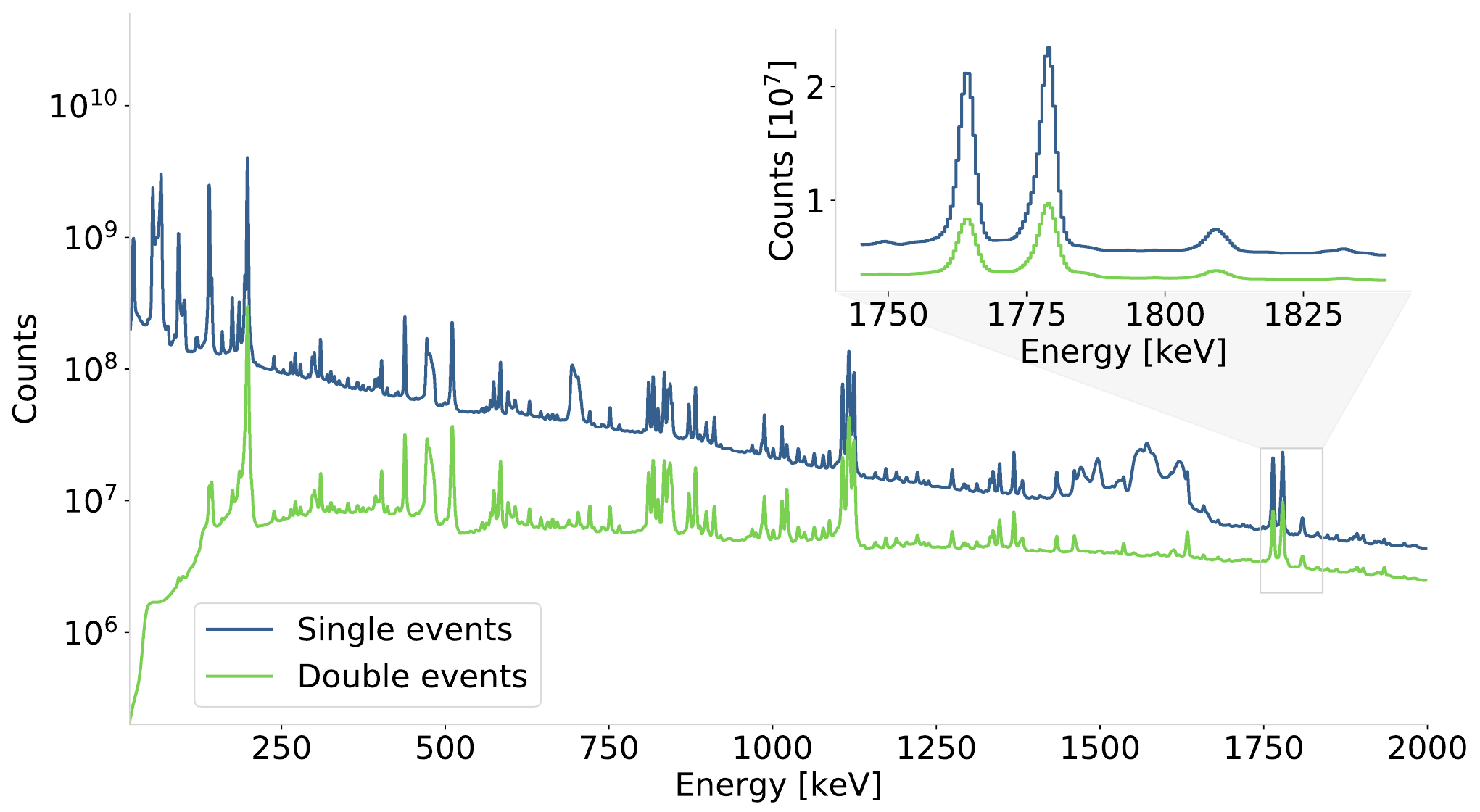}
\caption{The integrated energy spectrum of counts from the data used in this paper. Shown are single events (blue) and double events (green)
in the range 20--2000~keV; the inset enlarges the region around the \Al line used for astrophysical analysis here. }
\label{fig:rawSpectrum}       
\end{figure}

For each of these pointings, the data from single Ge detector hits, as well as those where two Ge detectors triggered within their 350~ns coincidence window (called "double events", DE) were binned into energy spectra for each of the detectors named 00-18; pairs of DE detectors are assigned detector ID's above 18, i.e., 19-60. Detectors are numerated from central toward outside in left-spiral counting, so ID=0 is the central detector and detectors 07-18 are on the outside of the hexagonal dense pack forming SPI's camera.
Event data from each detector are calibrated using known background lines, so that the distortions through gain variations are corrected for. Then these events are binned into spectra at 0.5~keV width, covering the 20--2000~keV range for response and background, and the 1790--1840~keV range for spectroscopy of the \Al line.
Figure~\ref{fig:rawSpectrum} shows the integrated spectra of these basic data, for all single events, and all double events, combined, respectively.
At the $^{26}$Al line energy of 1809 keV, the number of double-hit events corresponds to 56\% of the amount of single-hit events, all including instrumental background and the celestial $^{26}$Al signal.

\begin{figure}[ht] %%%%%%%%%%%%%%%%%%%%%%%%%%%%%%%%%%%%%%%%%
\centering
\includegraphics[width=0.8\columnwidth,clip]{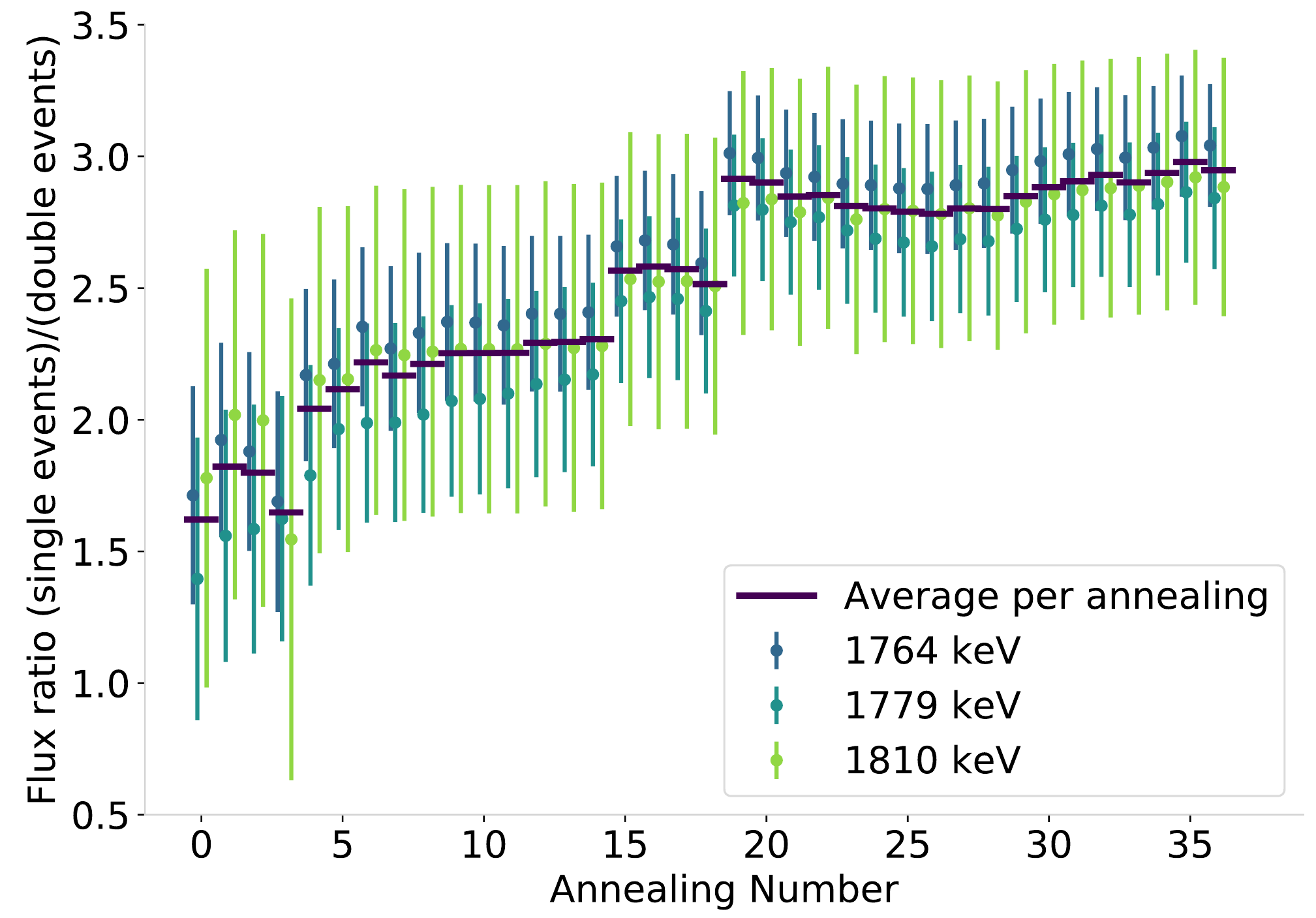}
\caption{The ratio of counts for single versus multiple events, over the time of the mission for the data used in this paper.
{Values integrated between} successive "detector annealings" {are shown}: these are performed roughly every { 6 months} to recover spectral resolutions after degradation due to cosmic-ray bombardment. {Detector failures occurred after 3$^{rd}$, 4$^{th}$, 12$^{th}$, and 15$^{th}$ annealings. Step changes after annealing 15 and 19 reflect  a reduction of the detector high voltage (4 to 3 to 2.5~kV) to help prolonged operations.}
}
\label{fig:ratioSE-ME}       
\end{figure}

In total, during the mission four of the nineteen Ge detectors failed, { with detector ID's 02/17/05/01 failing in} orbit {numbers 140/214/775/929, respectively.}. This means that { a different} camera response {applies after each failure,} for different numbers of operational detectors (19-18-17-16-15), {changing} in particular the rates of single versus double or multiple detector triggers (because failed detectors do not provide a 'multiple-event' trigger any more). 
Figure~\ref{fig:ratioSE-ME} shows how the ratio of single to multiple detectors undergoes these step-wise changes as detectors have failed over the course of the mission. In this Figure, a cumulative value is shown for the time between two annealing operations;  these are performed when cosmic-ray bombardment in orbit has degraded the charge collection properties of these semiconductor detectors up to a critical level, i.e. roughly every 6 months, through a 2-week heating period  \citep[for details see, e.g.][]{Diehl:2018}. 
This change in active detector number requires in total five different instrumental responses and background models, each for the respective camera configurations.

 %%%%%%%%%%%%%%%%%%%%%%%%%%%%%%%%%%%%%%%%%%%

 %%%%%%%%%%%%%%%%%%%%%%%%%%%%%%%%%%%%%%%%%%%

\subsection{Data Analysis}
\label{sec:analysis}

SPI data in the form of a set of detector count spectra for each spacecraft pointing and each detector make up our "data space". Per bin, we thus have  
$d_{i,j,k}$ measured counts, with $i,j,k$ being the indices of pointing, detector, and energy bin that span the data space.  
Here we not only use the spectra from the 19 physical detectors' events, but also spectra for double-hit events of adjacent Ge detectors; the respective detector combinations result in additional 42 "virtual" detectors, hence the detector variable $j\in[0,60]$ instead of $j\in[0,18]$.

These data are the result of the instrument's response to the $\gamma$-ray sky and the underlying instrumental background:
\begin{equation}
d_{i,j,k} = \sum_l R_{l;ijk} \sum_{n = 1}^{N_\mathrm{s}} \theta_n S_{nl} +
\sum_{n = N_\mathrm{s} + 1}^{N_\mathrm{s} + N_\mathrm{b}} \theta_n
B_{n;ijk}\label{eq:model-fit}
\end{equation}
Here, we identify sky model components $S_{n}$, such as point sources and diffuse emissions. These are formulated in "image space", as photon source intensities per sky direction $l$.  The instrument response matrix $R_{l;ijk}$ must be applied, to link the source locations on the sky to data, combining coordinate transformations per pointing to aspect angles, and then accounting for the individual mask/detector configurations of each pointing.
This sky signal is superimposed onto a large instrumental background, and we distinguish components e.g. from continuum and from lines reflecting specific processes. The background models are formulated in the same data space of detectors and their counts; no specific instrumental-response application is required, in particular no shadowing by the mask occurs, as background is recorded by the active detector volumes from all directions.
A comparison of the data as measured to predictions from models is obtained by maximum-likelihood fits of different model sets independently to each energy bin of our data space, thus obtaining the energy spectrum of intensities for the sky model, as a result.
	
Background is characterised by a continuum which falls off exponentially toward higher energies, and about 400 superimposed instrumental lines, reflecting the composition of spacecraft and instrument materials that are target to cosmic-ray interactions. \citep[Details can be found in][ where background and response are characterised in detail over the entire mission]{Diehl:2018}.  
The method of modeling instrumental background for the actually-used observations has been described before \citep{Siegert:2019}. 

\begin{figure}[ht] %%%%%%%%%%%%%%%%%%%%%%%%%%%%%%%%%%%%%%%%%
\centering
\includegraphics[width=0.8\columnwidth,clip]{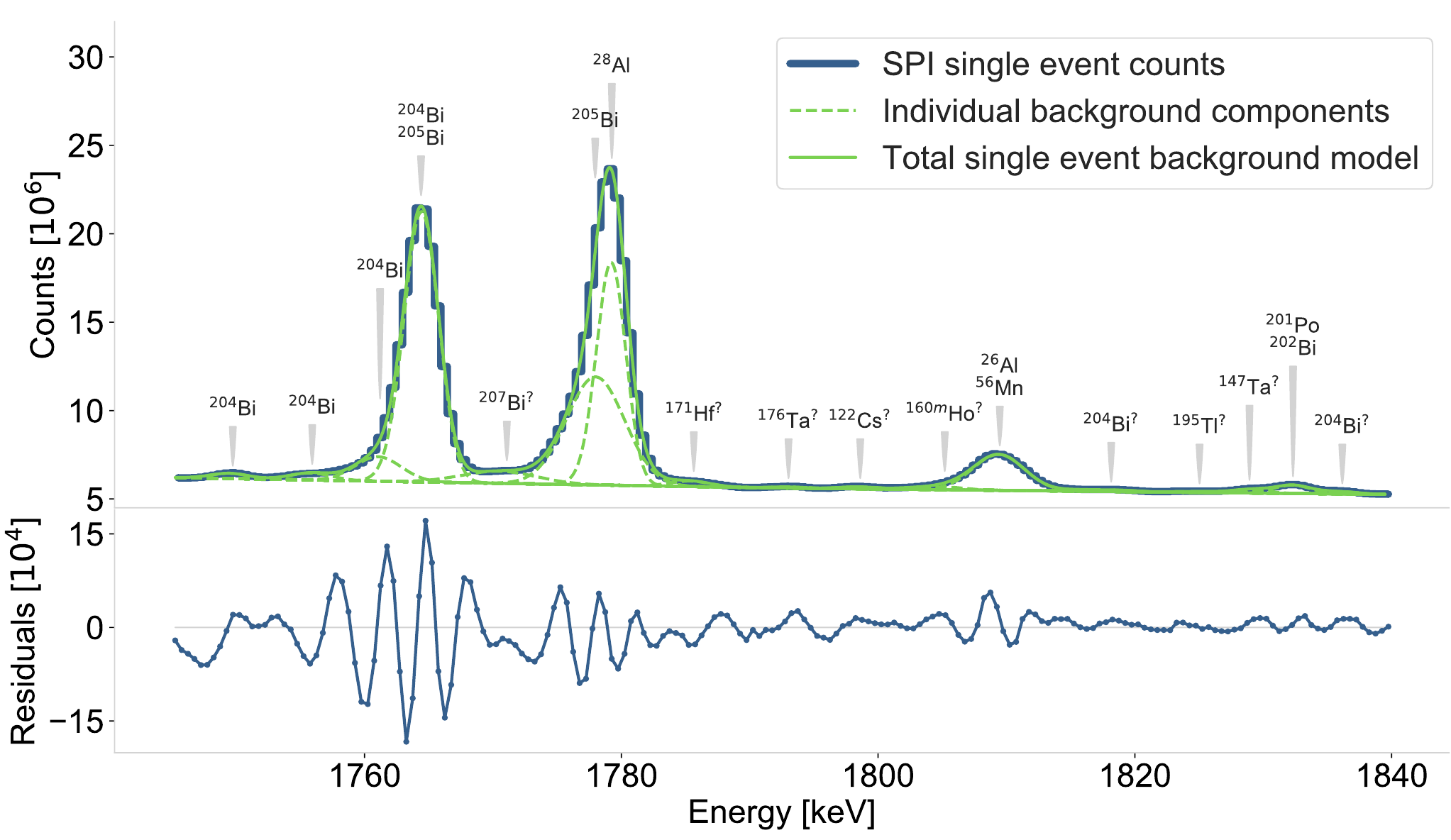}
\includegraphics[width=0.8\columnwidth,clip]{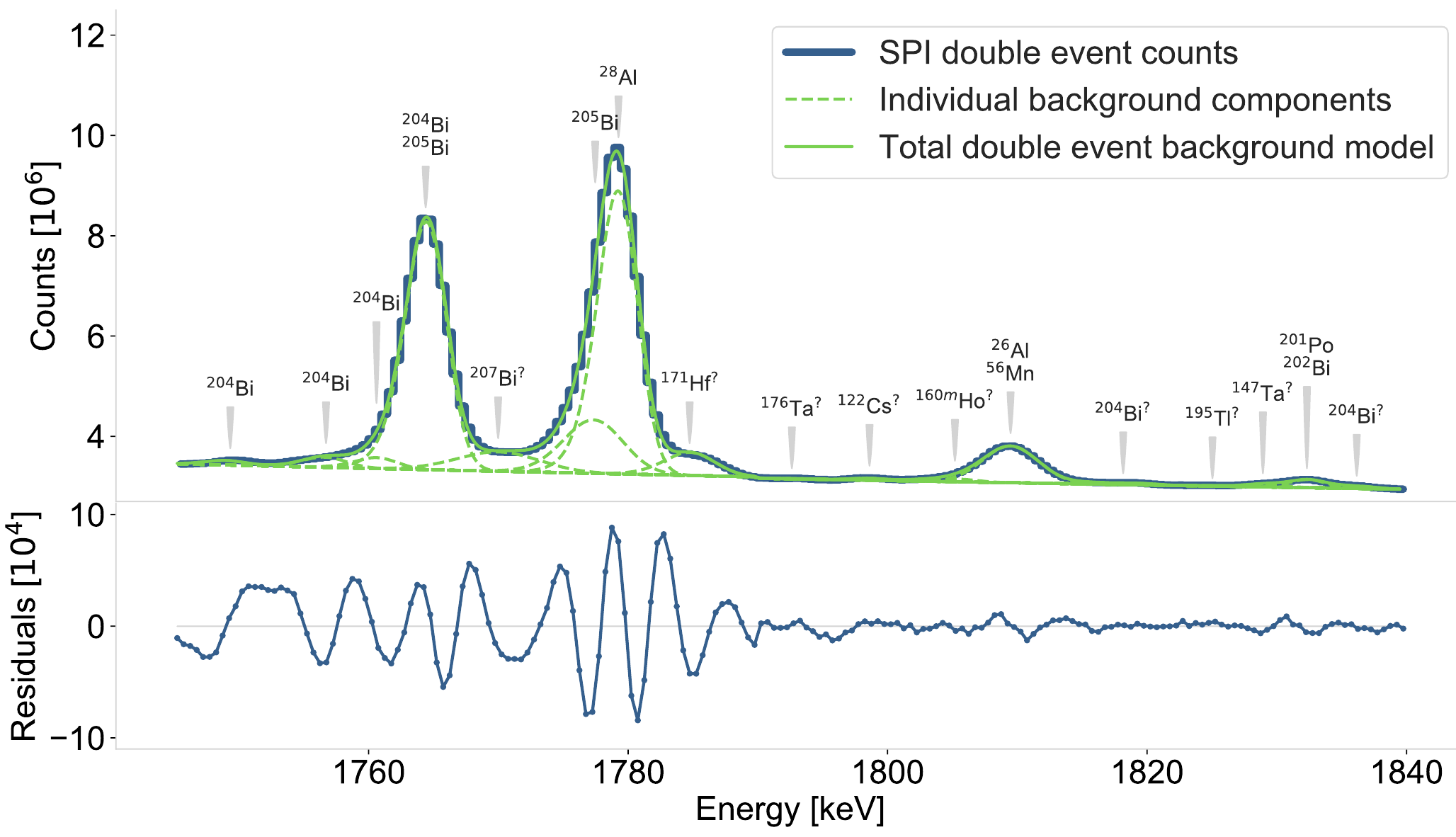}
\caption{The energy spectrum of the background model used for the single events (\emph{above}) and double events (\emph{below}).}
\label{fig:bgdModelSE}       
\end{figure} %%%%%%%%%%%%%%%%%%%%%%%%%%%%%%%%%%%%%%%%%%%

Figure~\ref{fig:bgdModelSE}  shows the performance of our background modeling in the energy region around the \Al line for single and double events, respectively, also displaying the spectrum of  residuals between actual data and the background model (lower graphs). Note that residuals are at the per-mille level in the regime of the \Al line, and the deviations from Gaussian line shapes (as assumed in our spectral model for the fits) result in residual structures at the percent level across strong background lines.
We note that the energy calibration and gain correction procedure as applied at the {\it INTEGRAL} Science Data Center, and the degradation from cosmic-ray bombardment with successive annealings, both lead to variations in line shapes that show up as such structured residuals across the regime of each strong background line. Additionally, many of the instrumental lines are blends among several lines with different intensities, as shown for the line near 1764~keV from different Bi isotopes.

\begin{figure}[ht] %%%%%%%%%%%%%%%%%%%%%%%%%%%%%%%%%%%%%%%%%
\centering
\includegraphics[width=0.65\columnwidth,clip]{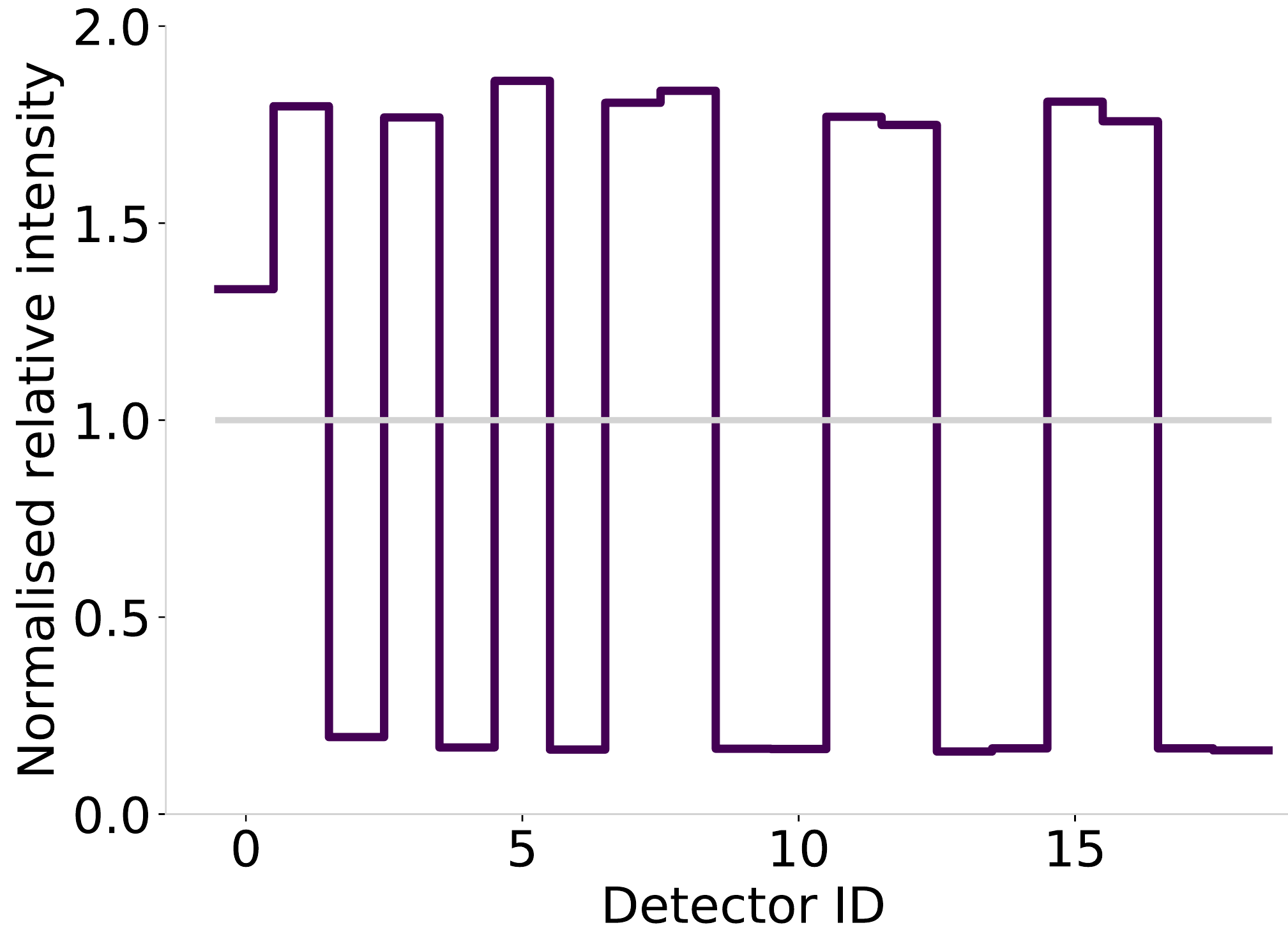}
\includegraphics[width=0.65\columnwidth,clip]{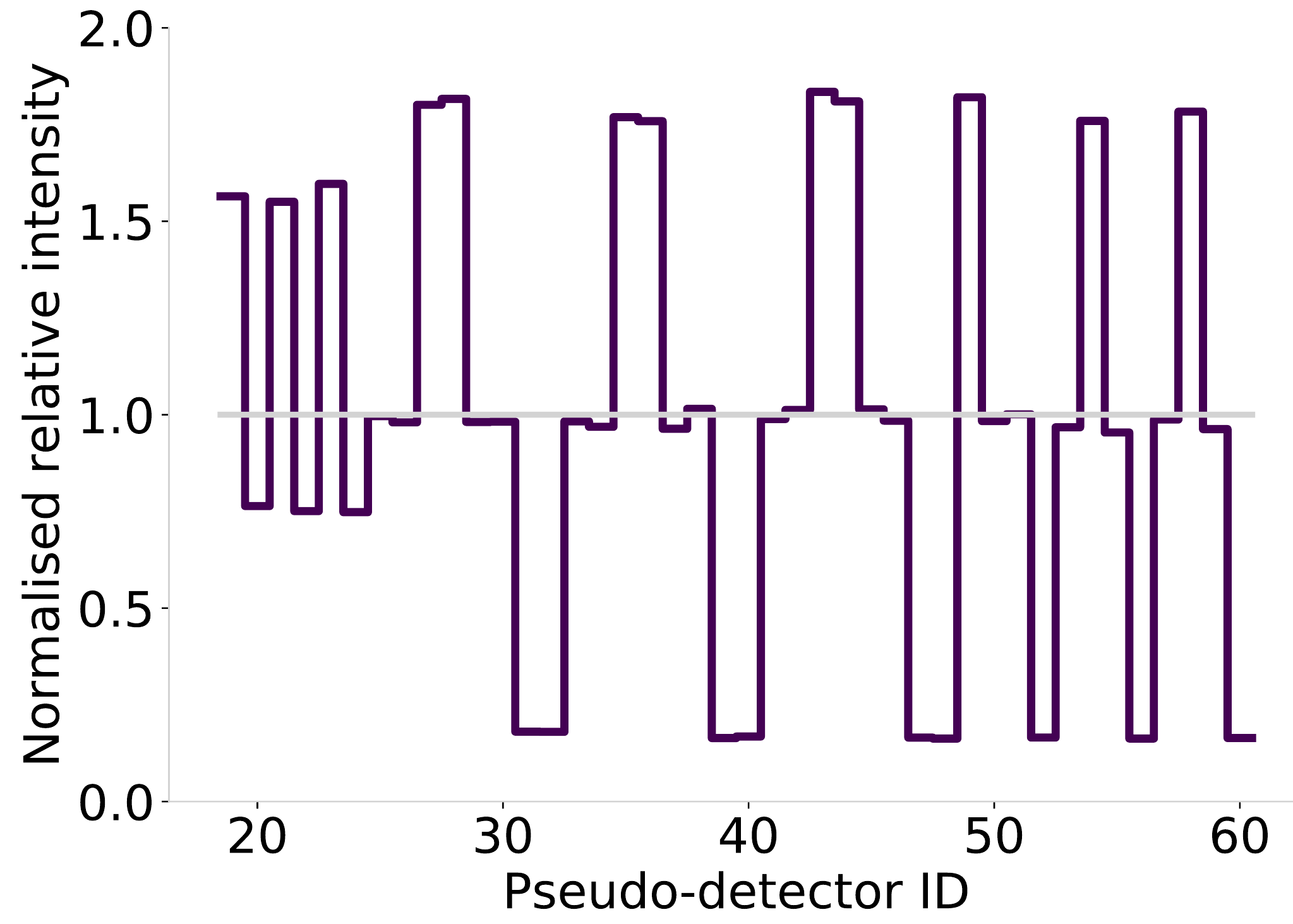}
\caption{The detector signature (ratio)  of a celestial source viewed on axis, for single and multiple events, respectively.}
\label{fig:detRatiosSrc}       
\end{figure}

The spectral response of SPI Ge detectors is characterized by a high photopeak efficiency, and a tail toward lower energies that arises from events where part of the primary photon energy is lost as a secondary photon escapes from the detector. 
Some of these escaping photons produce double events, and thus the full energy is recovered in the photopeak of double events. 
The flux per energy bin in the photopeak $i$ typically exceeds those in the Compton tail by at least one order of magnitude. Therefore, our spectral analysis typically only focuses on the photopeak, and adopts a basic Gaussian line shape. In order to account for spectral degradation between annealings, it has been found adequate to supplement this Gaussian by a one-sided exponential distribution extending downward in energy from the photopeak, fitting its width adding $\tau$ as a degradation parameter. 
This spectral model underlies the spectral fits of our instrumental lines \citep[see above and ][]{Diehl:2018}.
The differences in photopeak efficiencies among detectors generates the shadowgram signature for an on-axis direction, and is illustrated for the \Al line energy at 1.8~MeV by the histogram across detectors; we name this the "detector pattern", or "detector ratio", which is shown 
in Fig.~\ref{fig:detRatiosSrc} for single and double events, respectively. We normalise to 1.0 if a detector has $1/n$ of the total intensity, with $n$ the number of active detectors.

 %%%%%%%%%%%%%%%%%%%%%%%%%%%%%%%%%%%%%%%%%%%

\begin{figure}[ht] %%%%%%%%%%%%%%%%%%%%%%%%%%%%%%%%%%%%%%%%%
\centering
\includegraphics[width=0.65\columnwidth,clip]{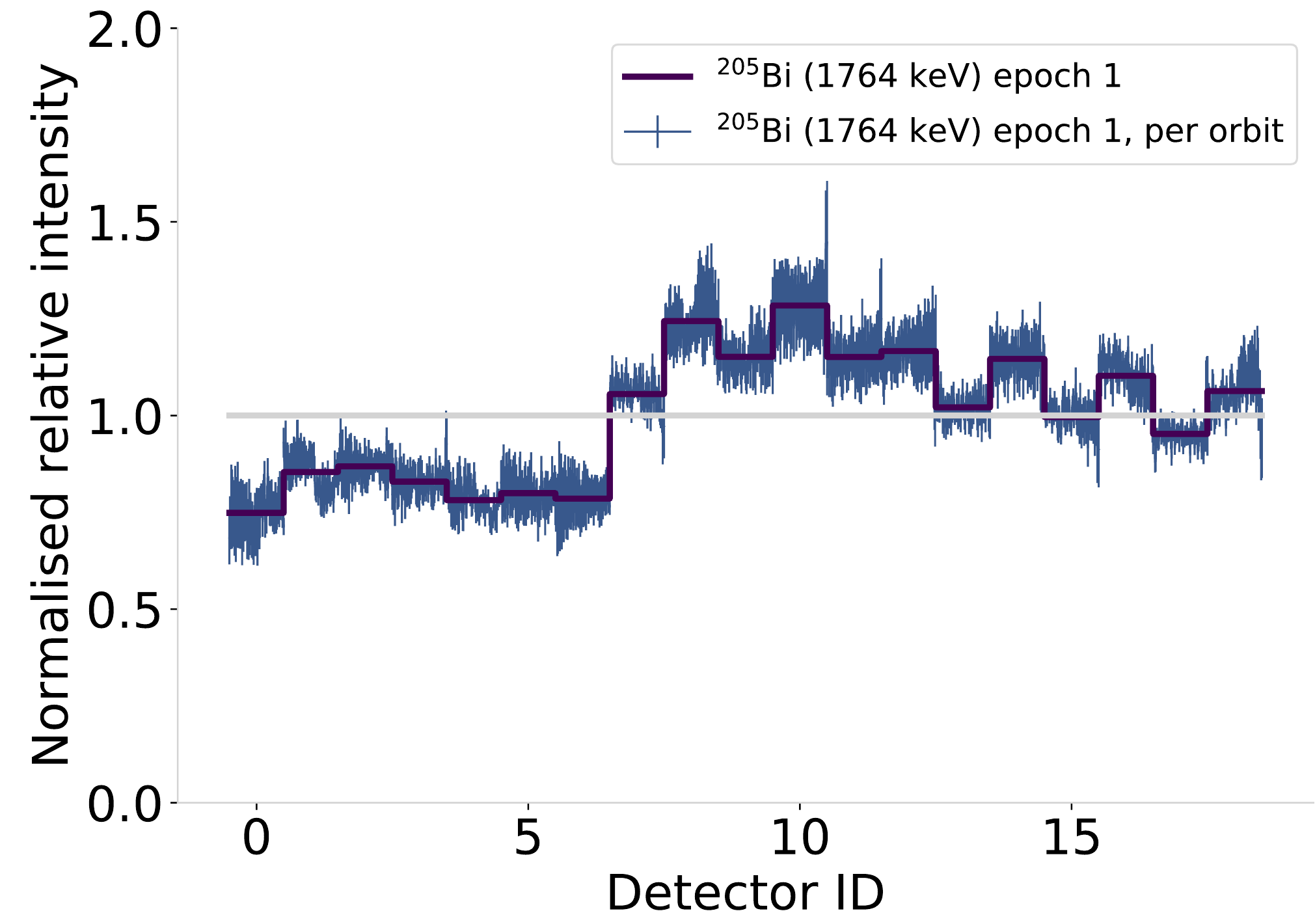}
\includegraphics[width=0.65\columnwidth,clip]{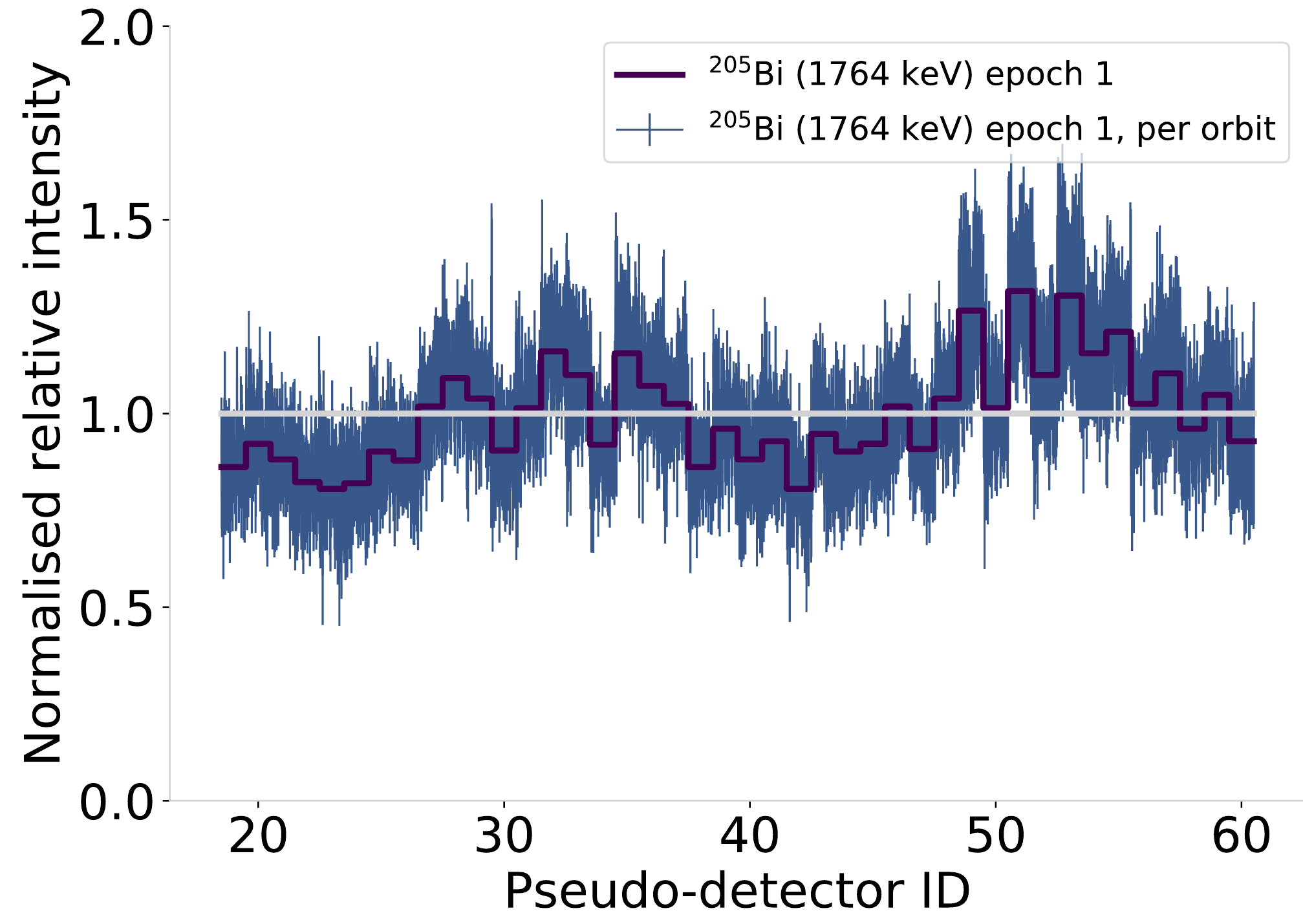}
\caption{The detector signature (ratio) of the background in the 1764 keV line, for single and multiple events, respectively. (See text for details).}
\label{fig:detRatios_bgd}       
\end{figure}

The discrimination of signals from the sky versus background occurs in a coded-mask instrument through the coding pattern implied by the mask above the instrument, which results in characteristic relative count distributions among detectors. 
In first approximation, instrumental background that mostly originates from cosmic-ray activated spacecraft materials should irradiate all detectors similarly; therefore, plausibly the corresponding {detector pattern} should be nearly flat for background.  
In detail, deviations and some structure in detector ratios occur, as some of the detectors may be more readily exposed to background from particular components of the spacecraft and instrument than others. For example, Ge background lines should be more intense in centrally-located detectors, which are surrounded by Ge everywhere, while Bi background lines should be less intense in central detectors, as outer detectors are exposed to the BGO detectors directly. 
The latter is illustrated in Fig.~\ref{fig:detRatios_bgd}, which shows the detector pattern for the background line near 1764~keV, which originates from different Bi isotopes as activated from cosmic rays. The upper graph shows results from single events, the lower from double events, for the first epoch of data where all 19 detectors still were operational. In fine scale along each detector ID, we superimpose the detector ratios for the different orbits within this epoch, to illustrate variations across shorter time scale. Evidently, the detector pattern remains rather stable, within few percent.
Each line, as well as the continuum, has its own characteristic detector pattern, which we apply in our background model, after it has been derived from the intensity of each Gaussian line fitted to each detector's spectra.  

%As 
{The} detector pattern for background remains {identical over time,} within a few percent { \citep{Diehl:2018}}, while the detector pattern for celestial gamma rays {(the "shadowgram")} varies as spacecraft pointings are {offset by 2$^\circ$}  on the 1/2 hour time scale. {We} thus obtain a sensitive discrimination of emission from the sky versus instrumental background.
In the maximum likelihood fit, we further allow adjustments of the absolute scale of the background with time, {accounting} for the fact that our primary data for background model determination already unavoidably include the emission from the sky. While this is {a local signature, and thus diluted to a} small {renormalization} effect {within statistical noise} for weaker point sources, the diffuse all-sky emission from \Al may produce detector patterns that are only weakly modulated and closer to those of background; thus {these} temporal {adjustments} of the relative {detector} ratios {in the background model improves its fit and}  helps { to discriminate also a weakly-modulated sky contribution}. We additionally benefit from our assumption that the \Al emission may vary with directions, but not in time.

 %%%%%%%%%%%%%%%%%%%%%%%%%%%%%%%%%%%%%%%%%%%

{Our fit maximizes the likelihood of model-predicted data in view of our measured data using Poissonian statistics, and optimizes model parameters $\theta$ therein to extract a spectrum of sky model coefficients $\theta_n(S)$, as described above.
In order to obtain a measure for the relative quality of such a fit, we derive a test statistic $\Phi$ that compares the log-likelihood difference between each model $M_1$ and the baseline hypothesis of fitting data $D$ with just the background model alone (hypothesis $M_0$):
$$ \Phi = 2 (log(L(D|M_1)-log(L(D|M_0)).$$
This allows a ranking/comparison of the performance for different models, higher values implying a better fit, low-end reference being the likelihood of the measured data if assuming background alone.}
{ It is important to check}  how the spectral result obtained from above analysis approach {depends} on the assumed morphology of \Al emission across the sky. For this purpose, we { typically} repeat our spectral fitting for a range of plausible emission morphologies, generated independently, revised{ or} from data that are unrelated to our analysis dataset itself. {In an earlier paper \citep{Wang:2009}, a detailed comparison of \Al line parameters for a set of plausible emission morphologies had been discussed. In particular Fig. 6 therein demonstrates that the centroid and width of the \Al line are practically independent of morphology for a wide range of plausible but different morphologies, while the line flux shows some variation among models. We concluded that the sensitivity of our spectral results to image morphology reduces to the total flux measured, and the line amplitude reflects how the sky model component matches SPI data in detail better than the background model alone. Therefore we now focus on an evaluation of line fluxes, and on how well different morphology models fit our dataset, after these have been convolved into our dataspace using the instrument response from coded mask and dithering.}

{Once we obtain a spectrum of sky model coefficients as obtained from likelihood optimisation of our fit, we extract spectral parameters through a Bayesian forward-folding approach: We describe the spectrum through a Gaussian line with flux $F$, centroid $E_0$, and width $\sigma$, plus a continuum approximated by a powerlaw function $c E^-{\alpha}$. 
Chosing appropriate priors for the uncertainty of each of these parameters, we then apply a Monte-Carlo-Markov-Chain (MCMC) fit to the extracted spectrum, forward folding this spectral model through the spectral response of SPI. MCMC analysis samples the posterior distribution for each parameter of the fit from given data and priors, maximising the full posterior distribution to determine best-fit parameter values. Our priors were based on the detailed spectral response analysis from data of the entire mission and energy range \citep{Diehl:2018}. 
The spectral response describes how incident photons are distributed among measured energies; it is dominated by the photopeak with instrumental line width $E_{SPI}$, and also includes the degradation parameter $\tau$, as described above. 
Note that, moreover, the Gaussian line width includes two components which we aim to separate for astrophysical interpretations: The instrumental line width $FWHM_{SPI}$ adds in quadrature to an astrophysical line broadening from the Doppler effect of \Al nuclei as they move in interstellar space while decaying, $FWHM_{sky}$. 
The posterior probability distributions for these 7 parameters that describe our spectral result are shown below, as is the probability distribution of the photon model itself (green area in spectra shows the 95\% confidence region of the model).}

\subsection{Models for Spatial Distribution of the Emission}
\label{sec:skymodels}

For our baseline analysis, we used the image obtained from the COMPTEL 9-year allsky survey, as it had been produced through maximum entropy deconvolution \citep{Diehl:1995b,Pluschke:2001c}.%; this result is shown in Figure~\ref{fig:skymapCOMPTEL}.
This map is thought to best represent our knowledge of \Al emission from the Galaxy based on observations \citep[see][for a discussion]{Prantzos:1996a,Diehl:1995i,Knodlseder:1999}.
We could use the image obtained for {\it INTEGRAL}/SPI with likelihood analysis \citep{Bouchet:2015}, which was obtained with background modeling from a general detector pattern and including a single normalisation parameter for each pointing, and free parameters of many pixels on the sky. Using this result would imply using partly overlapping data, and thus is not pursued here.

Alternatively, we {could} employ models taken from astrophysical considerations of the spatial distribution of \Al sources in our Galaxy, or from tracers that plausibly may represent these \citep[e.g.][]{Diehl:1997h,Knodlseder:1999a}.
{This had been pursued in detail for comparison of \Al and \Fe models \citep[see][]{Wang:2020}, and thus is not repeated here.}

As another alternative {pursued more recently}, we have constructed a bottom-up model for a galaxy, starting from models of stars with their evolution and nucleosynthesis yields, building stellar groups using a mass spectrum, and finally placing these groups into a spatial model for the galaxy. Drawing random realisations of such a model by sampling the parameters included in such a bottom-up stars-groups-galaxy model, one obtains a stellar population synthesis based model of a galaxy. We have used the PSYCO implementation of such an approach {\citep[see details in][]{Pleintinger:2020,Siegert:2022}}.
This provides us with a predictive model for the appearance of the observed \Al sky, evaluating the Galaxy's content of \Al from theory inputs concerning source evolution and nucleosynthesis output, allowing us to compare observed flux to expectations from theory.	
%----------------------------------------------------------------------------------------------------------

\section{Results}
\label{sec:results}

\begin{figure}[ht] %%%%%%%%%%%%%%%%%%%%%%%%%%%%%%%%%%%%%%%%%
\centering
\includegraphics[width=0.9\columnwidth,clip]{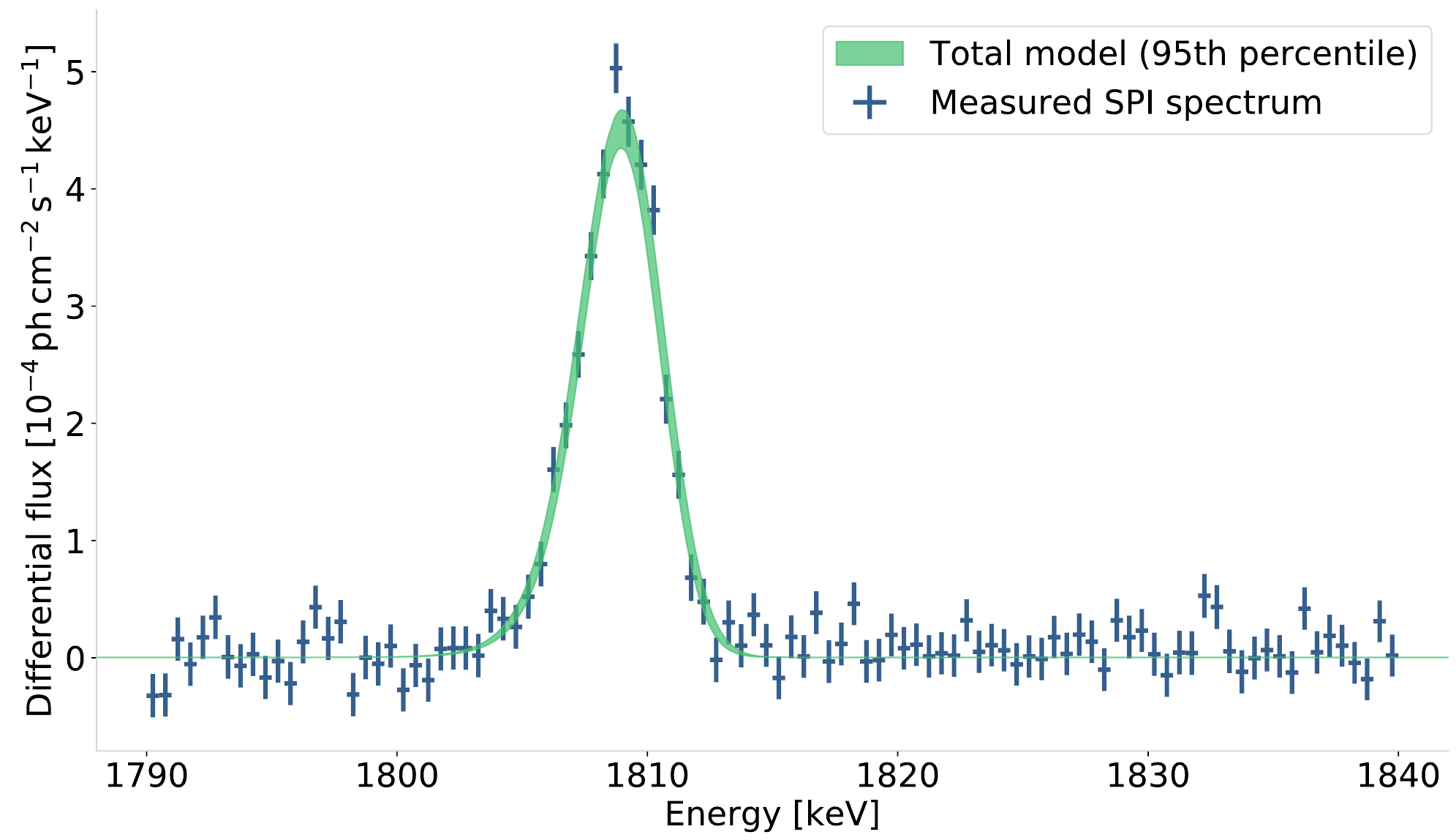}
\includegraphics[width=0.9\columnwidth,clip]{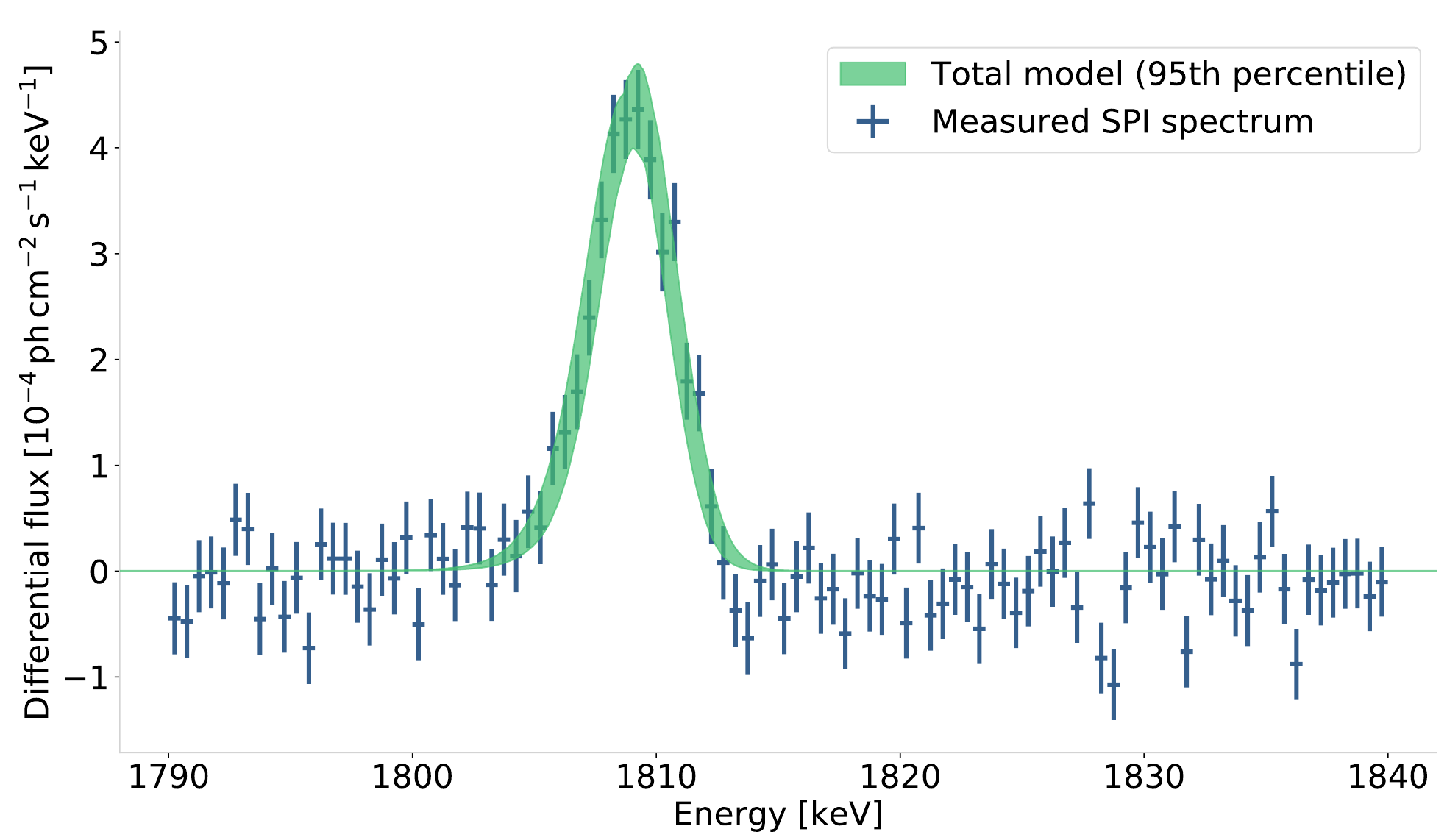}
\includegraphics[width=0.9\columnwidth,clip]{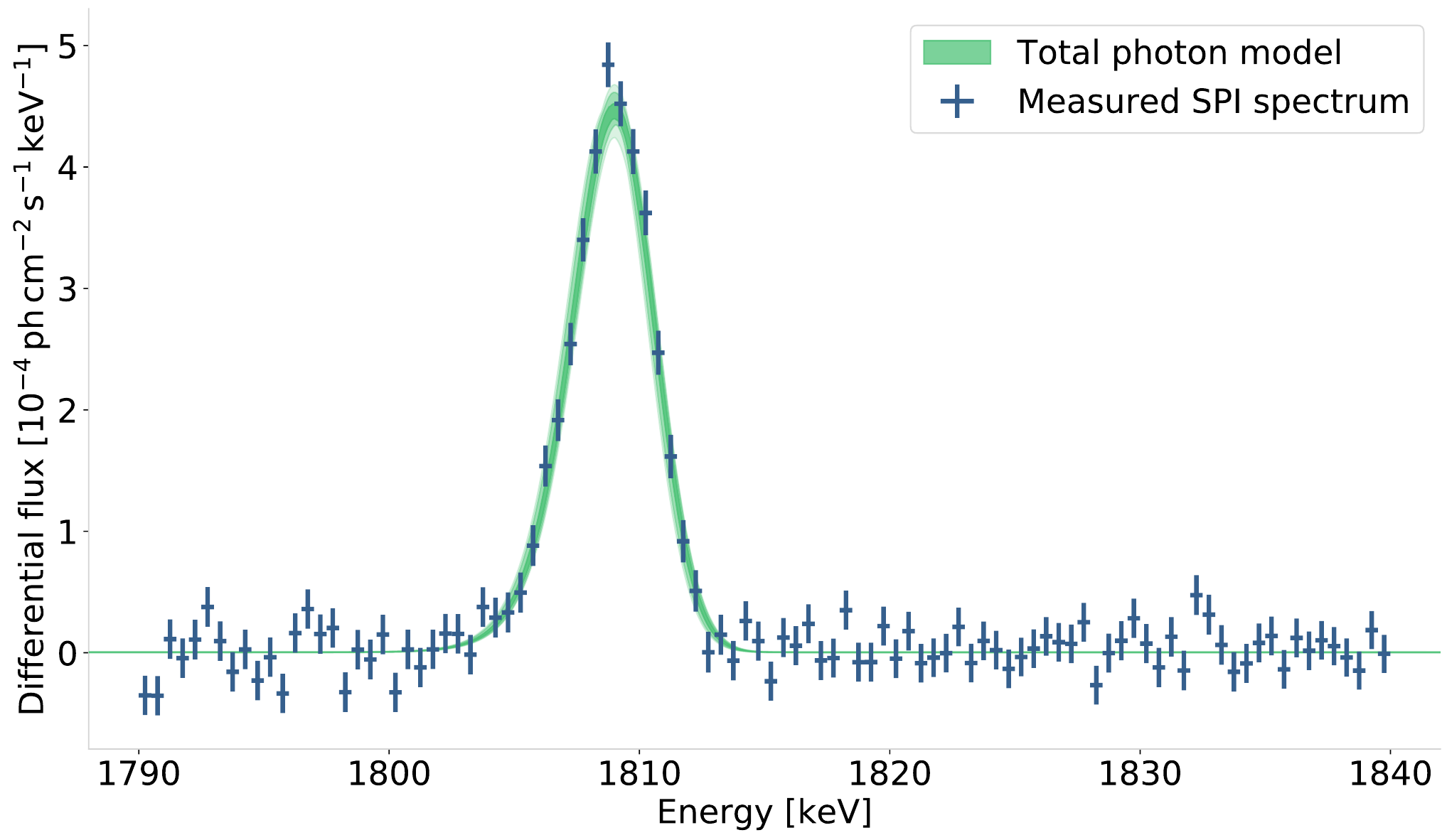}
\caption{The energy spectrum around the $^{26}$Al line, as obtained from single-hit events (top), double-hit events (middle), and combined events (bottom) of our observations.
Data points represent the measurement, with bin width and Poissonian statistical uncertainty indicated. The uncertainties shown in shaded green have been derived from MCMC modeling of the spectral response, and show variations {within 95\% of}  the model.}
\label{fig:specAllsky_SE-ME}       
\end{figure}

\begin{figure}[ht] %%%%%%%%%%%%%%%%%%%%%%%%%%%%%%%%%%%%%%%%%
\centering
\includegraphics[width=1.0\columnwidth,clip]{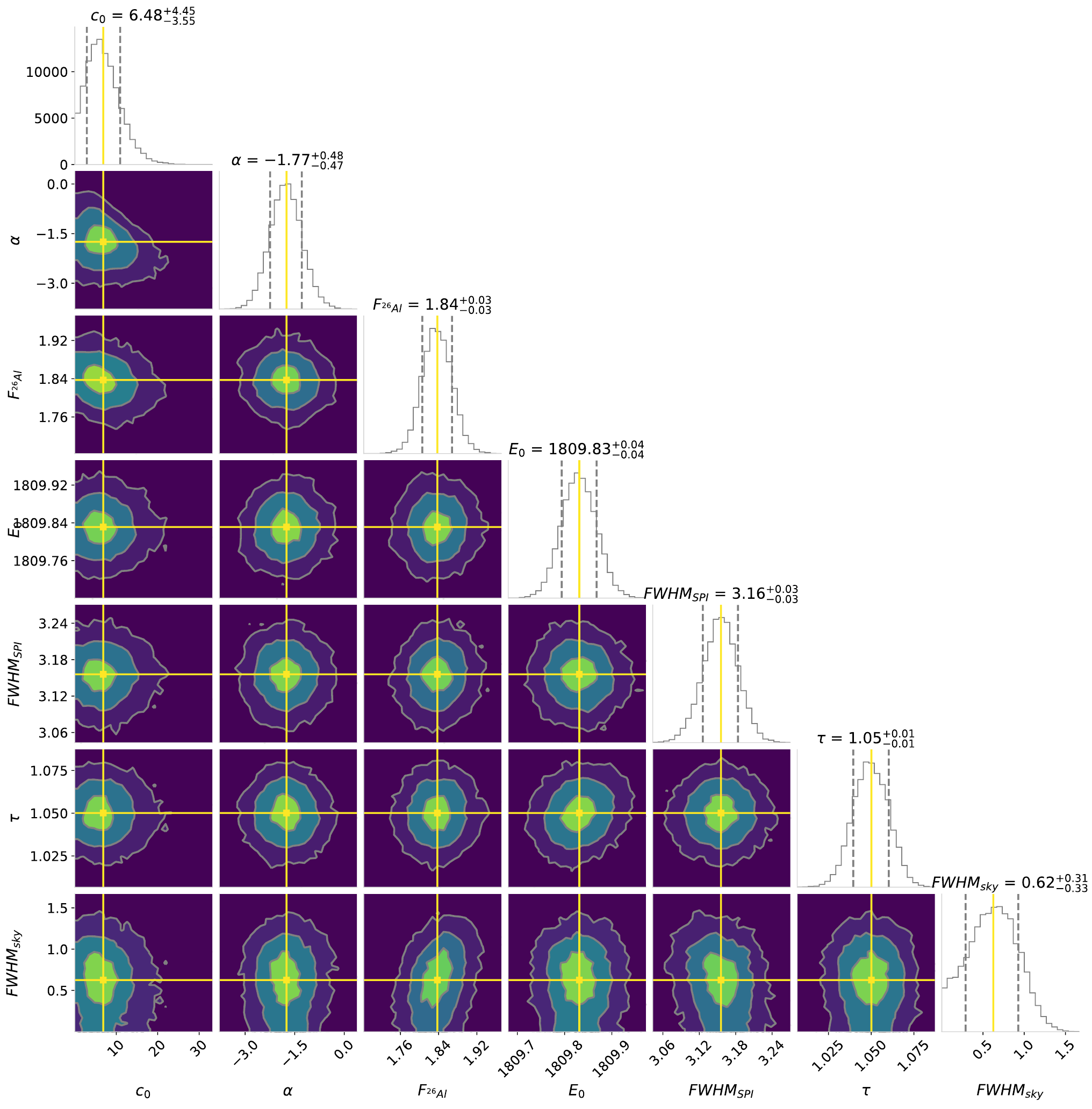}
\caption{The {posterior distributions of spectral }parameters  around the $^{26}$Al line {of the total-events spectrum,} Fig.~\ref{fig:specAllsky_SE-ME} (bottom), illustrating confidence regions {for each parameter}.}
\label{fig:specParametersAllsky}       
\end{figure}

Using 17.5 years of data and the COMPTEL-derived sky distribution of \Al emission, we derive spectra around the \Al line at 1808.65~keV shown in Fig.~\ref{fig:specAllsky_SE-ME} for single and for double events, separately, and combining single and double events.
Spectral parameter values and their uncertainties (confidence regions) have been derived from MCMC analysis and are shown in Fig.~\ref{fig:specParametersAllsky}, illustrating the uncertainties of the fit.  
Fitting background models separately per event type, but a single set of line characteristics, a total detection significance of 58$\sigma$ is found for the \Al line. Single-hit events only show a significance of 51$\sigma$, and double-hit events alone provide a clear 27$\sigma$ detection above background.
Table~\ref{tab:linefits} provides fit results for single and double-hit events, { and for data combining both event types.}
Intensity, line centroid, and line width are consistent within uncertainties between single-hit, double hit, {and combined-event} results.
{Note that the 'total' result is derived through a separate analysis chain (see Section 2.2), rather than averaging single and double event results.} 

We find that the spectral response appears to be less sharp for double events, as spectral resolution of SPI for single events is 3.17\,keV at 1.8\,MeV \citep{Diehl:2018}.
This is attributed to a bias in locations of events within Ge detectors, double events favoring events detected from interactions in outer regions, rather than closer to the central anode. Charge collection properties vary across the detector with radius from the anode, and a bias in location will plausibly produce a line shape deviating from the all-detector sample in single events (which was confirmed to be nearly Gaussian after annealings).

Decomposing the line width into contributions from instrumental resolution as measured with other instrumental lines versus line broadening of the celestial contribution, we obtain a consistent celestial broadening of  (0.46$\pm$0.3)~keV and  (0.76$\pm$0.6)~keV, respectively for single and double events. This leads to a combined result of  (0.62$\pm$0.03)~keV, which again is dominated by the single-events contribution.

The line centroid is found at (1809.83$\pm$0.04)~keV from combined data. The laboratory value is 1808.65(7)~keV, % 1808.70(6) \citep{Endt:1990}. now cited from Wu & Browne LBNL 1997 in ENSDF2020
  and thus the integrated $^{26}$Al emission from the Galaxy appears blue-shifted by about 1.2~keV.
The all-sky flux for the $^{26}$Al line is determined as (1.84$\pm$0.03$)\times$10$^{-3}$~ph~cm$^{-2}$s$^{-1}$. 

 %%%%%%%%%%%%%%%%%%%%%%%%%%%%%%%%%%%%%%%%%%%

 %%%%%%%%%%%%%%%%%%%%%%%%%%%%%%%%%%%%%%%%%%%

\begin{table} %%%%%%%%%%%%%%%%%%%%%%%%%%%%%%%%%%%%%%
		\caption{Fitted Spectral Line Parameters.  Intensity is given in units $\left[ 10^{-3} \frac{\text{ph}}{\text{cm}^2 \text{s}} \right]$, centroid and width (FWHM) are in keV units.}             
\label{tab:linefits}      
\centering          
\begin{tabular}{|l | c | c | c |}
			\hline\hline       
			& single hits &  double hits & total\\ \hline
	 intensity  &  1.83$\pm$0.03                 & 1.79$\pm$0.06              & 1.84$\pm$0.03  \\
	 centroid                                                & 1809.80 $\pm$0.04           &  1809.91 $\pm$0.1        & 1809.84 $\pm$0.04  \\
	 width                                                  & 3.20 $\pm$0.05                 & 3.49 $\pm$0.17            & 3.23 $\pm$0.07 \\
			\hline                  
		\end{tabular}
\end{table} %%%%%%%%%%%%%%%%%%%%%%%%%%

\begin{figure}[ht] %%%%%%%%%%%%%%%%%%%%%%%%%%%%%%%%%%%%%%%%%
\centering
\includegraphics[width=1.0\columnwidth,clip]{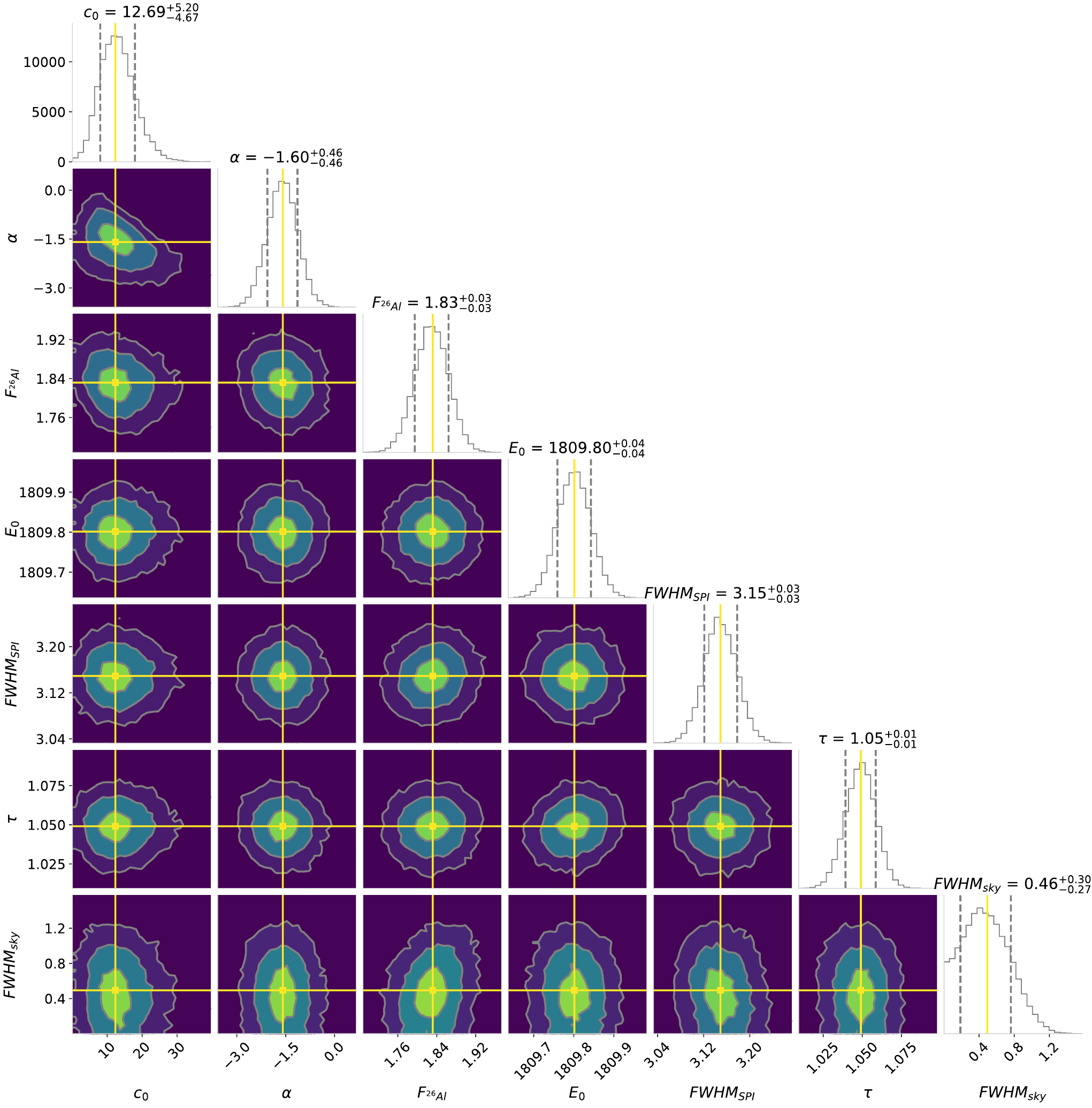}
\includegraphics[width=1.0\columnwidth,clip]{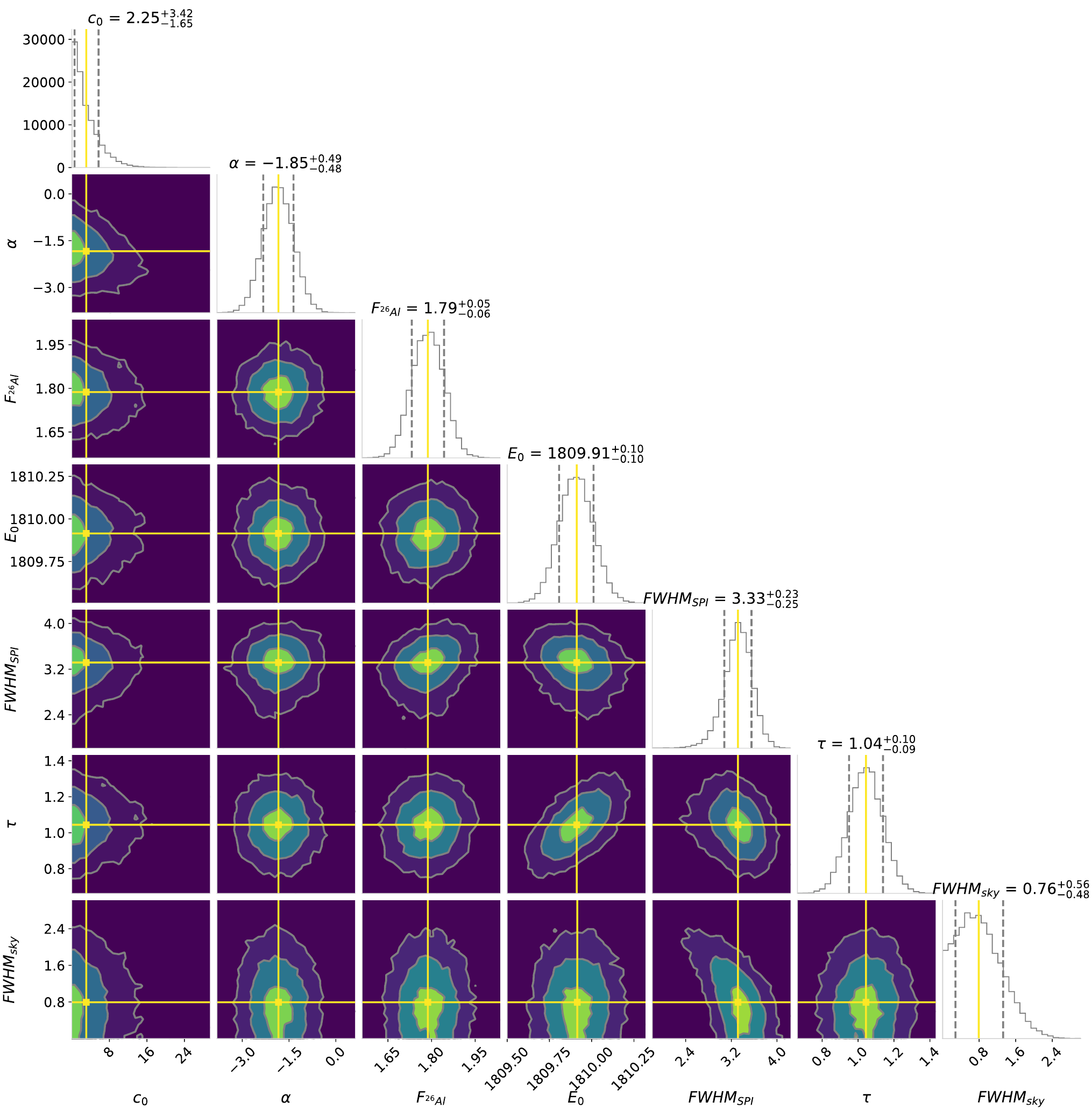}
\caption{{Posterior distributions of spectral parameters  around the $^{26}$Al line for the spectra derived separately for single (top) and double events (bottom).}}
\label{fig:specParametersAllsky_eventtypes}       
\end{figure} %%%%%%%%%%%%%%%%%%%%%%%%%%%%%%%%%%%%%%%%%%%

{Our baseline model for the emission morphology, taken from the COMPTEL \Al results, obtains a test statistic value of 2160.
We cannot use the emission morphology derived from  SPI for a scientific result as it uses partially-overlapping data; but still this provides an interesting reference for our test statistic,  obtaining $\Phi=$2166 for such a self-consistent image morphology.
By comparison, our best-fitting bottom-up population synthesis model PSYCO obtains a test statistic value $\Phi$=~2061, while other PSYCO variants discussed in detail in companion paper \citep{Siegert:2022} obtain $\Phi$ values between 1500 and 2061.}

\section{Discussion}
\label{sec:discussion}

Our dataset comprises single-hit and double-hit events; at the $^{26}$Al line energy of 1809 keV, the number of double-hit events corresponds to 56\% of the amount of single-hit events, all including instrumental background and the celestial $^{26}$Al signal.
Applying our analysis of fitting models of instrumental background and sky, and including spectral response details, this results in an increase of the signal from celestial $^{26}$Al through the additional use of double-hit events that would be equivalent to an additional observing time of 30\%.
 
The main characteristics of $^{26}$Al emission from the Galaxy as found from earlier work are confirmed by our analysis including also double-hit detector events.
 The line centroid at (1809.83$\pm$0.04~keV implies a blue shift corresponding to Doppler velocities of order 100 km~s$^{-1}$.
 We attribute this to our aspect of emission regions throughout the Galaxy, which, due to large-scale Galactic rotation, are moving at such typical velocities with respect to the local standard of rest \citep[see][for longitude-resolved analysis]{Kretschmer:2013}. 
   Line broadening from source kinematics integrated over the sky is found to be  (0.62$\pm0.3$)~keV (FWHM). This is consistent with the Doppler broadening from regions at different bulk velocities to the observer, as incurred from large-scale Galactic rotation and the dispersal of \Al into superbubbles, as found before \citep{Kretschmer:2013}, {with implications discussed, e.g. in \citet{Krause:2015}}.

We find an all-sky flux of (1.84$\pm$0.03$)\times$10$^{-3}$~ph~cm$^{-1}$cm$^{-1}$  for the $^{26}$Al line from our analysis of 18.5~years of data and using single as well as double events. Note that uncertainties now are at the level of 2\% (see Fig.~\ref{fig:specParametersAllsky}). 

The flux attributed to the "inner Galaxy" is $\sim$3$ \times$10$^{-4}$~ph~cm$^{-2}$s$^{-1}$, or 16\% of the total flux.
The inner Galaxy has often been  used for comparisons, and may be defined as a longitude range $\pm$30 degrees around the Galactic centre, or, slightly differently, the "inner radian" (57 degrees).  This region often had been analysed as a reference toward obtaining results representative for the Galaxy as a whole, thus focusing on the brightest region and avoiding issues from low surface-brightness contributions that often are subject to systematic uncertainties from background determinations. Different latitude ranges have been used, as well, from $\pm$5 to 15 degrees.
   We note that the majority of the integrated flux is found outside the {inner Galaxy}, as reported earlier \citep[see][and table 1 herein]{Pleintinger:2019}. This is in tension with simulations and expectations, which report typically 50\% of the all-sky flux coming from the inner Galaxy. Special nearby regions, such as Cygnus, Scorpius-Centaurus, and Orion, have been known before to potentially lead to distortions of such Galaxy-wide conclusions. It remains a challenge to properly account for flux contributions from nearby regions, in particular as the imaging performance of the coded mask degrades for diffuse emission that extends over large parts of the entire sky.
   
An earlier analysis  \citep{Wang:2020} found an all-sky \Al flux of  (1.68$\pm$0.07$)\times$10$^{-3}$~ph~cm$^{-2}$s$^{-1}$. They used geometrical models only, for the spatial distribution of the emission, and this flux was derived for a best-matching double-exponential disk size with scale parameters radius=7~kpc and height=800~pc. 
Another earlier analysis using 13.5~years of data \citep{Pleintinger:2019} found an all-sky flux of (1.71$\pm$0.06)$\times$10$^{-3}$~ph~cm$^{-2}$s$^{-1}$ from using single event hits only. Here, the COMPTEL map had been used as a model for spatial distribution. 
Using different spatial models for $^{26}$Al emission morphology, flux results typically vary within $\leq$10\% \citep[see, e.g.][]{Diehl:2010}.
These cases illustrate that the overall flux of the sky in \Al $\gamma$~rays is rather well constrained, and our flux determined from using single and double events agrees with these, within uncertainties. 

We note that our bottom-up population synthesis model PSYCO predicts a substantially-lower flux value \citep[see][for details and discussion]{Siegert:2022}.
Recently, from the {\it COSI} balloon flight of 2016, a flux measurement in the \Al line was published \citep{Beechert:2022} that is somewhat (2$\sigma$)  higher than our measurement: 
they find (8.6$\pm$2.5$)\times$10$^{-4}$~ph~cm$^{-2}$s$^{-1}$ from the 'inner Galaxy', which compares to $\sim$3$\times$10$^{-4}$~ph~cm$^{-2}$s$^{-1}$ from our PSYCO-map based analysis. They discuss potential calibration issues for {\it COSI} absolute efficiency. Nevertheless, it is important to perform measurements with instruments that carry different systematics, such as different fields of view and imaging methods.

%----------------------------------------------------------------------------------------------------------

\section{Summary and Conclusions}
\label{sec:conclusions}

  Using {\it INTEGRAL} SPI events of types single- and double-hits, we enhance the {detection efficiency} for Galactic $^{26}$Al emission. 
  We apply improved spectral response and background as evaluated from tracing all-event spectral details over the entire mission, and re-determine the intensity of Galactic $^{26}$Al emission across the entire sky{. Herein we apply} maximum likelihood fits of simulated and model-built sky distributions to { measurements of single and double events in {\it INTEGRAL's} SPI detectors.} 
  Spectral results are consistent with previous measurements, within uncertainties.
   A shift in the line centroid with respect to the laboratory value reflects an integrated blue-shift attributed to large-scale Galactic rotation and aspect of regions with their respective relative velocities to the solar system.
      Line broadening confirms the Doppler broadening from regions at different bulk velocities to the observer, as incurred from large-scale Galactic rotation and the dispersal of \Al into superbubbles {\citep{Kretschmer:2013,Krause:2015}}.
         We find an all-sky flux of (1.84$\pm$0.03$)\times$10$^{-3}$~ph~cm$^{-1}$cm$^{-1}$ that appears somewhat larger than upscaling {from values for} the inner Galaxy {\citep{Wang:2009,Prantzos:1996}.  This indicates that \Al} emission {also extends to regions beyond the Galactic plane and toward} higher latitudes. {Such emission is} attributed to nearby superbubbles, {that relate to the nearby Scorpius-Centaurus groups of massive stars \citep{Krause:2018}, the local bubble \citep{Zucker:2022}, and chimneys that connect large superbubbles with the Galactic halo  \citep{Pleintinger:2019,Krause:2021}. Large-scale Galactic \Al emission has been used to infer, e.g., the total  \Al mass in our Galaxy and the supernova rate \citep[e.g.][]{Diehl:2006d}, and have been based on inner-Galaxy flux values. Although their conclusions largely remain identical, these are subject to small adjustments  from improved precision of  \Al flux values in regions that properly characterize our Galaxy at large.}   
%%%%%%%%%%%%%%%%%
\section*{Acknowledgments}
This study was supported by the Deutsche Forschungsgemeinschaft (DFG, German Research Foundation) under its Excellence Strategy, the Munich Clusters of Excellence "Origin and Structure of the Universe" and "Origins" (EXC-2094-390783311), and by the EU through COST action ChETEC CA160117.
  The {\it INTEGRAL}/SPI project
  has been completed under the responsibility and leadership of CNES;
  we are grateful to ASI, CEA, CNES, DLR, ESA, INTA, NASA and OSTC for
  support of this ESA space science mission.

%----------------------------------------------------------------------------------------------------------
	
\bibliographystyle{aa}
%\bibliography{rod-refs18}

\begin{thebibliography}{29}
\expandafter\ifx\csname natexlab\endcsname\relax\def\natexlab#1{#1}\fi

\bibitem[{{Beechert} {et~al.}(2022){Beechert}, {Siegert}, {Tomsick},
  {Zoglauer}, {Boggs}, {Brandt}, {Gulick}, {Jean}, {Kierans}, {Lazar},
  {Lowell}, {Roberts}, {Sleator}, \& {von Ballmoos}}]{Beechert:2022}
{Beechert}, J., {Siegert}, T., {Tomsick}, J.~A., {et~al.} 2022, \apj, 928, 119

\bibitem[{{Bouchet} {et~al.}(2015){Bouchet}, {Jourdain}, \&
  {Roques}}]{Bouchet:2015}
{Bouchet}, L., {Jourdain}, E., \& {Roques}, J.-P. 2015, \apj, 801, 142

\bibitem[{{Diehl}(1995)}]{Diehl:1995i}
{Diehl}, R. 1995, Experimental Astronomy, 6, 103

\bibitem[{{Diehl} {et~al.}(1995){Diehl}, {Dupraz}, {Bennett}, {Bloemen},
  {Hermsen}, {Knoedlseder}, {Lichti}, {Morris}, {Ryan}, {Schoenfelder},
  {Steinle}, {Strong}, {Swanenburg}, {Varendorff}, \& {Winkler}}]{Diehl:1995b}
{Diehl}, R., {Dupraz}, C., {Bennett}, K., {et~al.} 1995, \aap, 298, 445

\bibitem[{{Diehl} {et~al.}(2006){Diehl}, {Halloin}, {Kretschmer}, {Lichti},
  {Sch{\"o}nfelder}, {Strong}, {von Kienlin}, {Wang}, {Jean}, {Kn{\"o}dlseder},
  {Roques}, {Weidenspointner}, {Schanne}, {Hartmann}, {Winkler}, \&
  {Wunderer}}]{Diehl:2006d}
{Diehl}, R., {Halloin}, H., {Kretschmer}, K., {et~al.} 2006, \nat, 439, 45

\bibitem[{{Diehl} {et~al.}(2003){Diehl}, {Kretschmer}, {Pl{\"u}schke},
  {Sch{\"o}nfelder}, {Strong}, {Cervi{\~n}o}, \& {Hartmann}}]{Diehl:2003}
{Diehl}, R., {Kretschmer}, K., {Pl{\"u}schke}, S., {et~al.} 2003, Astronomische
  Nachrichten Supplement, 324, 18

\bibitem[{{Diehl} {et~al.}(2010){Diehl}, {Lang}, {Martin}, {Ohlendorf},
  {Preibisch}, {Voss}, {Jean}, {Roques}, {von Ballmoos}, \&
  {Wang}}]{Diehl:2010}
{Diehl}, R., {Lang}, M.~G., {Martin}, P., {et~al.} 2010, \aap, 522, A51+

\bibitem[{{Diehl} {et~al.}(1997){Diehl}, {Oberlack}, {Kn{\"o}dlseder},
  {Bloemen}, {Hermsen}, {Morris}, {Ryan}, {Sch{\"o}nfelder}, {Strong}, {von
  Ballmoos}, \& {Winkler}}]{Diehl:1997h}
{Diehl}, R., {Oberlack}, U., {Kn{\"o}dlseder}, J., {et~al.} 1997, in American
  Institute of Physics Conference Series, Vol. 410, Proceedings of the Fourth
  Compton Symposium, ed. C.~D. {Dermer}, M.~S. {Strickman}, \& J.~D. {Kurfess},
  1114--1118

\bibitem[{{Diehl} {et~al.}(2018){Diehl}, {Siegert}, {Greiner}, {Krause},
  {Kretschmer}, {Lang}, {Pleintinger}, {Strong}, {Weinberger}, \&
  {Zhang}}]{Diehl:2018}
{Diehl}, R., {Siegert}, T., {Greiner}, J., {et~al.} 2018, \aap, 611, A12

\bibitem[{{Kn{\"o}dlseder} {et~al.}(1999{\natexlab{a}}){Kn{\"o}dlseder},
  {Bennett}, {Bloemen}, {Diehl}, {Hermsen}, {Oberlack}, {Ryan},
  {Sch{\"o}nfelder}, \& {von Ballmoos}}]{Knodlseder:1999a}
{Kn{\"o}dlseder}, J., {Bennett}, K., {Bloemen}, H., {et~al.}
  1999{\natexlab{a}}, \aap, 344, 68

\bibitem[{{Kn{\"o}dlseder} {et~al.}(1999{\natexlab{b}}){Kn{\"o}dlseder},
  {Dixon}, {Bennett}, {Bloemen}, {Diehl}, {Hermsen}, {Oberlack}, {Ryan},
  {Sch{\"o}nfelder}, \& {von Ballmoos}}]{Knodlseder:1999}
{Kn{\"o}dlseder}, J., {Dixon}, D., {Bennett}, K., {et~al.} 1999{\natexlab{b}},
  \aap, 345, 813

\bibitem[{{Krause} {et~al.}(2018){Krause}, {Burkert}, {Diehl}, {Fierlinger},
  {Gaczkowski}, {Kroell}, {Ngoumou}, {Roccatagliata}, {Siegert}, \&
  {Preibisch}}]{Krause:2018}
{Krause}, M. G.~H., {Burkert}, A., {Diehl}, R., {et~al.} 2018, \aap, 619, A120

\bibitem[{{Krause} {et~al.}(2015){Krause}, {Diehl}, {Bagetakos}, {Brinks},
  {Burkert}, {Gerhard}, {Greiner}, {Kretschmer}, \& {Siegert}}]{Krause:2015}
{Krause}, M.~G.~H., {Diehl}, R., {Bagetakos}, Y., {et~al.} 2015, \aap, 578,
  A113

\bibitem[{{Krause} {et~al.}(2021){Krause}, {Rodgers-Lee}, {Dale}, {Diehl}, \&
  {Kobayashi}}]{Krause:2021}
{Krause}, M. G.~H., {Rodgers-Lee}, D., {Dale}, J.~E., {Diehl}, R., \&
  {Kobayashi}, C. 2021, \mnras, 501, 210

\bibitem[{{Kretschmer} {et~al.}(2013){Kretschmer}, {Diehl}, {Krause},
  {Burkert}, {Fierlinger}, {Gerhard}, {Greiner}, \& {Wang}}]{Kretschmer:2013}
{Kretschmer}, K., {Diehl}, R., {Krause}, M., {et~al.} 2013, \aap, 559, A99

\bibitem[{{Pleintinger}(2020)}]{Pleintinger:2020}
{Pleintinger}, M. M.~M. 2020, PhD thesis, Technische Universit{\"a}t
  M{\"u}nchen

\bibitem[{{Pleintinger} {et~al.}(2019){Pleintinger}, {Siegert}, {Diehl},
  {Fujimoto}, {Greiner}, {Krause}, \& {Krumholz}}]{Pleintinger:2019}
{Pleintinger}, M. M.~M., {Siegert}, T., {Diehl}, R., {et~al.} 2019, \aap, 632,
  A73

\bibitem[{{Pl{\"u}schke} {et~al.}(2001){Pl{\"u}schke}, {Diehl},
  {Sch{\"o}nfelder}, {Bloemen}, {Hermsen}, {Bennett}, {Winkler}, {McConnell},
  {Ryan}, {Oberlack}, \& {Kn{\"o}dlseder}}]{Pluschke:2001c}
{Pl{\"u}schke}, S., {Diehl}, R., {Sch{\"o}nfelder}, V., {et~al.} 2001, in ESA
  Special Publication, Vol. 459, Exploring the Gamma-Ray Universe, ed.
  A.~{Gimenez}, V.~{Reglero}, \& C.~{Winkler}, 55--58

\bibitem[{{Prantzos}(1996)}]{Prantzos:1996}
{Prantzos}, N. 1996, \aaps, 120, C303+

\bibitem[{{Prantzos} \& {Diehl}(1996)}]{Prantzos:1996a}
{Prantzos}, N. \& {Diehl}, R. 1996, \physrep, 267, 1

\bibitem[{{Roques} {et~al.}(2003){Roques}, {Schanne}, {von Kienlin},
  {Kn{\"o}dlseder}, {Briet}, {Bouchet}, {Paul}, {Boggs}, {Caraveo},
  {Cass{\'e}}, {Cordier}, {Diehl}, {Durouchoux}, {Jean}, {Leleux}, {Lichti},
  {Mandrou}, {Matteson}, {Sanchez}, {Sch{\"o}nfelder}, {Skinner}, {Strong},
  {Teegarden}, {Vedrenne}, {von Ballmoos}, \& {Wunderer}}]{Roques:2003}
{Roques}, J.~P., {Schanne}, S., {von Kienlin}, A., {et~al.} 2003, \aap, 411,
  L91

\bibitem[{{Siegert} {et~al.}(2022){Siegert}, {Pleintinger}, {Diehl}, {Krause}, {Greiner},
  \& {Weinberger}}]{Siegert:2022}
{Siegert}, T., {Pleintinger}, M., {Diehl}, R., {Krause}, M., {Greiner}, J., \& {Weinberger},
  C. 2022, submitted to \aap

\bibitem[{{Siegert} {et~al.}(2019){Siegert}, {Diehl}, {Weinberger},
  {Pleintinger}, {Greiner}, \& {Zhang}}]{Siegert:2019}
{Siegert}, T., {Diehl}, R., {Weinberger}, C., {et~al.} 2019, \aap, 626, A73
 
\bibitem[{{Siegert} {et~al.}(2016){Siegert}, {Diehl}, {Khachatryan}, {Krause},
  {Guglielmetti}, {Greiner}, {Strong}, \& {Zhang}}]{Siegert:2016}
{Siegert}, T., {Diehl}, R., {Khachatryan}, G., {et~al.} 2016, \aap, 586, A84

\bibitem[{{Vedrenne} {et~al.}(2003){Vedrenne}, {Roques}, {Sch{\"o}nfelder},
  {Mandrou}, {Lichti}, {von Kienlin}, {Cordier}, {Schanne}, {Kn{\"o}dlseder},
  {Skinner}, {Jean}, {Sanchez}, {Caraveo}, {Teegarden}, {von Ballmoos},
  {Bouchet}, {Paul}, {Matteson}, {Boggs}, {Wunderer}, {Leleux},
  {Weidenspointner}, {Durouchoux}, {Diehl}, {Strong}, {Cass{\'e}}, {Clair}, \&
  {Andr{\'e}}}]{Vedrenne:2003}
{Vedrenne}, G., {Roques}, J.-P., {Sch{\"o}nfelder}, V., {et~al.} 2003, \aap,
  411, L63

\bibitem[{{Wang} {et~al.}(2009){Wang}, {Lang}, {Diehl}, {Halloin}, {Jean},
  {Kn{\"o}dlseder}, {Kretschmer}, {Martin}, {Roques}, {Strong}, {Winkler}, \&
  {Zhang}}]{Wang:2009}
{Wang}, W., {Lang}, M.~G., {Diehl}, R., {et~al.} 2009, \aap, 496, 713

\bibitem[{{Wang} {et~al.}(2020){Wang}, {Siegert}, {Dai}, {Diehl}, {Greiner},
  {Heger}, {Krause}, {Lang}, {Pleintinger}, \& {Zhang}}]{Wang:2020}
{Wang}, W., {Siegert}, T., {Dai}, Z.~G., {et~al.} 2020, \apj, 889, 169

\bibitem[{{Winkler} {et~al.}(2003){Winkler}, {Courvoisier}, {Di Cocco},
  {Gehrels}, {Gim{\'e}nez}, {Grebenev}, {Hermsen}, {Mas-Hesse}, {Lebrun},
  {Lund}, {Palumbo}, {Paul}, {Roques}, {Schnopper}, {Sch{\"o}nfelder},
  {Sunyaev}, {Teegarden}, {Ubertini}, {Vedrenne}, \& {Dean}}]{Winkler:2003}
{Winkler}, C., {Courvoisier}, T.~J.-L., {Di Cocco}, G., {et~al.} 2003, \aap,
  411, L1

\bibitem[{{Zucker} {et~al.}(2022){Zucker}, {Goodman}, {Alves}, {Bialy},
  {Foley}, {Speagle}, {Gro{\^I}{\texttwosuperior}schedl}, {Finkbeiner},
  {Burkert}, {Khimey}, \& {Swiggum}}]{Zucker:2022}
{Zucker}, C., {Goodman}, A.~A., {Alves}, J., {et~al.} 2022, \nat, 601, 334

\end{thebibliography}

\end{document}